\documentclass[a4paper]{article}

\usepackage{INTERSPEECH2021}
\usepackage{multirow}

\title{GIST-AiTeR System for the Diarization Task of the 2022 VoxCeleb Speaker Recognition Challenge}
\name{
    Dongkeon Park$^1$,
    Yechan Yu$^2$, 
    Kyeong Wan Park$^2$,
    Ji Won Kim$^1$,
    and Hong Kook Kim$^{1,2}$
    }
\address{
    $^1$AI Graduate School,
    $^2$School of Electrical Engineering and Computer Science\\
    Gwangju Institute of Science and Technology, Gwangju 61005, Republic of Korea
}
\email{\{dongkeon@, yechan1202@gm., pakinlab4194@gm., jiwon.kim@gm., hongkook@\}gist.ac.kr}

\begin{document}

\maketitle

\begin{abstract} 
    This report describes the submission system of the GIST-AiTeR team at the 2022 VoxCeleb Speaker Recognition Challenge (VoxSRC) Track 4.
Our system mainly includes speech enhancement, voice activity detection , multi-scaled speaker embedding, probabilistic linear discriminant analysis-based speaker clustering, and  overlapped speech detection models.
We first construct four different diarization systems according to different model combinations with the best experimental efforts.
Our final submission is an ensemble system of all the four systems and achieves a diarization error rate of 5.12\% on the challenge evaluation set, ranked third at the diarization track of the challenge.

\end{abstract}

\noindent\textbf{Index Terms}: VoxSRC-22, Speaker Diarization, Multi-scale embedding, Overlapped Speech Detection, VoxConverse

\section{Introduction} 
    Speaker diarization (SD) is a process of identifying “who spoke when” by dividing an audio signal into homogeneous segments using speaker labels \cite{tranter2003investigation}.
SD is essential to many speech-related applications with multi-speaker audio data, such as conversational multi-part speech recognition for business meetings or interviews and speaker-dependent video indexing \cite{reynolds05approaches}.

In general, SD has been considered a speaker clustering problem that  assigns a speaker label to each speech segment.
A clustering-based SD system typically has a modular structure, comprising speech activity detection, a speaker  embedding extractor, and speaker clustering \cite{Shum2013}.
After a speaker label is assigned  to each segment, all segments with the same speaker label are grouped into a cluster.
As the speaker embedding becomes increasingly  robust, the conventional SD system also achieves good performance because  the speaker confusion is significantly reduced.
To further improve the performance, many research works have focused  on overlapped speech detection (OSD) \cite{overlap4518619} to reduce the missed speaker error, including speech separation  \cite{xiao2021microsoft}, end-to-end neural speaker diarization (EEND) \cite{fujita2020end},  and target-speaker voice activity detection (TS-VAD)  \cite{medennikov2020target}.

In this report, we propose a clustering-based SD system for the Diarization Task of the 2022 VoxCeleb Speaker Recognition Challenge (VoxSRC).
The proposed system comprises several processing modules such as speech enhancement, voice activity detection (VAD), speaker embedding extraction, clustering, and OSD.

Following this introduction, Section 2 describes the proposed system architecture, where we develop four different systems according to the different combinations of SD  modules with different hyperparameters.
In addition, the fusion systems are described to improve the diarization performance.
Further, Section 3 explains the datasets used for the challenge, and Section 4 discusses the experimental results.
Finally, Section 5 concludes  this report.
\label{sec:intro}
    \begin{figure}[!t]
	\centering
	\includegraphics[width=0.49\textwidth]{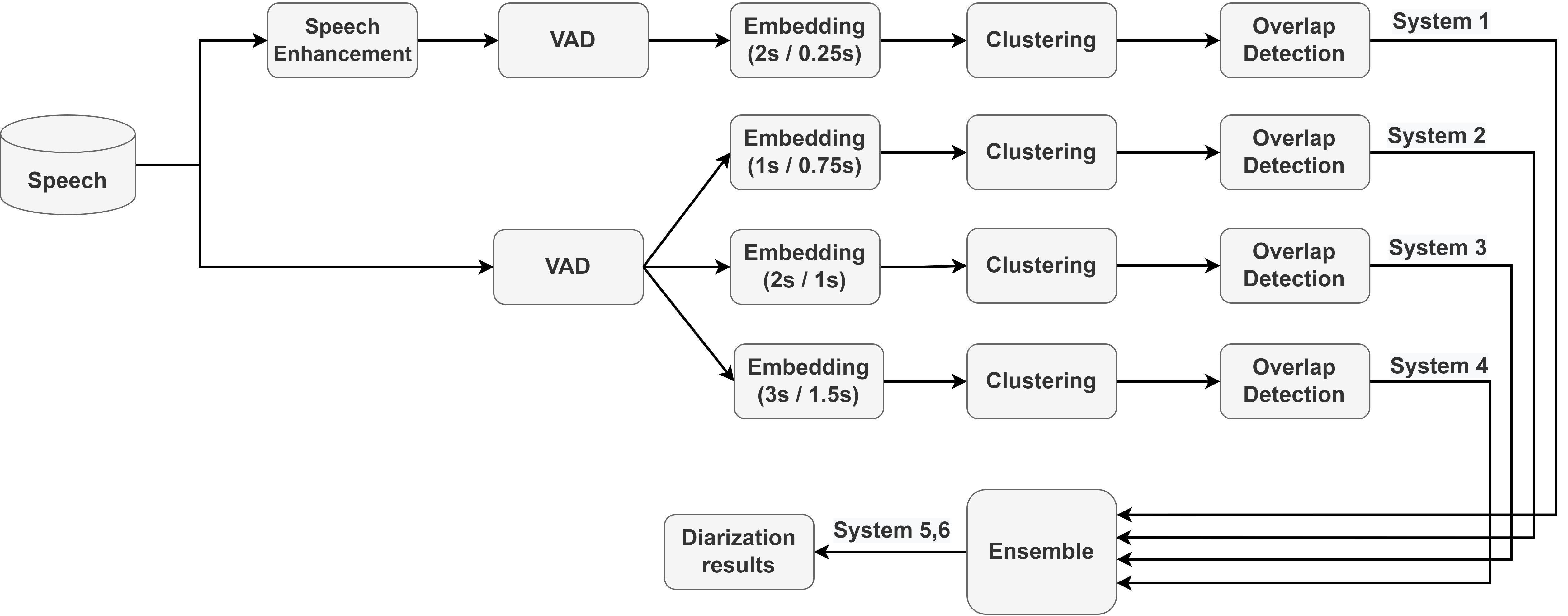}
	\caption{
    	Overall procedure of the GIST-AiTeR systems for the diarization task, where System 6 was finally submitted. 
    }
	\label{fig:arch}
\end{figure}

\section{GIST-AiTeR Speaker Diarization System}\label{sec:system}
    Figure \ref{fig:arch} illustrates the overall procedure of our developed SD systems. The following subsections will explain each SD module in detail.
    
    \subsection{Speech Enhancement}\label{sec:se} 
        As a preprocessing method, FullSubNet-based speech enhancement was used \cite{hao2021fullsubnet}, which was already proposed in the Deep Noise Suppression (DNS) Challenge \cite{reddy2020interspeech} organized by INTERSPEECH 2020 for single-channel speech enhancement.
The FullSubNet comprises a full-band model and a sub-band model.
The full-band model can capture the global spectral information and extract the cross-band dependencies, while the sub-band model attends to the local spectral pattern.
We plug a pretrained FullSubNet to our SD system\footnote{https://github.com/haoxiangsnr/FullSubNet}.

    \subsection{Voice Activity Detection}\label{sec:vad}
        VAD is the first step of all the SD systems, as shown in Fig. \ref{tab:vad}.
We constructed two VAD modules of ResNet+LSTM and SincNet+LSTM and compared them in terms of VAD and SD performances.
Because no ground-truth VAD labels were available in this challenge, diarization labels were used as VAD labels for training VAD.


The ResNet+LSTM module is identical to the model proposed in \cite{wang2021dku}, except for the front-end model and statistical pooling size $S$.
In other words, we used ResNetSE \cite{hu2018squeeze} as a front-end model for the global property modeling instead of ResNet34 \cite{he2016deep}.
To train ResNet+LSTM, we uniformly chunked both the mixed training set and DEV402 into segments of  8 s long with a step size of 4 s.
Then, a sequence of 80-dimensional log mel-filterbanks were obtained by applying a Hamming window of 25 ms to each segment once every 10 ms, followed by a 512-point fast Fourier transform.
The model was first trained using the mixed training set for 10 epochs with a learning rate of 0.0001 and a weight decaying factor of 1e-5 for preventing overfitting of the model, where binary cross-entropy (BCE) loss and Adam optimizer were used.
Subsequently, the model was fine-tuned using speech-enhanced DEV402 for 10 epochs with a learning rate of 0.00001.
In particular, we trained two different versions of ResNet+LSTM with different $S = \{1, 2\}$ to obtain the VAD outputs with different resolutions.
Finally, we averaged the outputs from these two versions in a frame level.
The threshold for VAD was experimentally set to 0.6.

As the second VAD module, we first took the pretrained model of SincNet+LSTM that was learned using the Pyannote 2.0 framework \cite{bredin2021end}.
The pretrained model was designed to provide three subtasks: VAD, speaker change detection, and OSD.
We only considered the VAD subtask to perform transfer learning using the DEV402 set without speech enhancement.

Finally, we constructed two ensemble modules by averaging the posterior values predicted from the ResNet+LSTM and SincNet+LSTM modules.
The ensemble values were postprocessed according the procedure in Pyannote2.0 \cite{bredin2021end}.
During the postprocessing, we optimized hyperparameters by applying grid search to VAL46 to achieve better performance.

Table \ref{tab:vad} compares the VAD performances of ResNet+LSTM, SincNet+LSTM, and their different ensemble modules by measuring the false alarm (FA), miss detection (Miss), and accuracy (Acc) on VAL46.
As presented in the table, Fusion (1+2+3) showed the best VAD performance.
We evaluated diarization error rate (DER) performance after each VAD module was integrated into the overall SD system.
Accordingly, an SD system employing SincNet+LSTM was found to achieve the best DER performance; therefore, SincNet+LSTM was selected as a VAD module in our final SD system.

\begin{table}[t]
  \caption{Comparison of the false alarm (FA), miss detection (Miss) and accuracy (Acc) of three different VAD models and their fusions.}
  \label{tab:vad}
  \centering
  \begin{tabular}[c]{lccc}
    \toprule
     & \textbf{FA (\%)} & \textbf{Miss (\%)} & \textbf{Acc} (\%) \\
    \midrule
    1. ResNet+LSTM (S = 1)  & 2.03    & 1.61  & 96.34  \\
    2. ResNet+LSTM (S = 2)  & 2.39    & 1.47  & 96.14  \\
    3. SincNet+LSTM      & 2.23    & 1.47  & 96.30  \\
    Fusion (1+2)         & 2.16    & 1.54  & 96.31  \\
    Fusion (1+2+3)       & 2.08    & 1.51  & 96.41  \\
    
    \bottomrule
  \end{tabular}
\end{table}

    \subsection{Speaker Embedding}\label{sec:xvec}
        A speaker embedding module is based on MFA-Conformer \cite{zhang2022mfa}.
Because of its performance in speaker recognition with a reasonable inference speed, MFA-Conformer was adopted to train the speaker embedding model.
To train MFA-Conformer, we used voxceleb1 \cite{Nagrani17} and voxceleb2 \cite{Chung18b}.
As an input feature, 80-dimensional log mel-filterbanks were used, which were identical to the input feature of the ResNet+LSTM VAD module.
Additionally, SpecAugment \cite{park2019specaugment} ($T$=8, $F$=10) was applied to these input features.
To realize multi-scaled input features, each utterance was randomly cropped into 1, 2, 3 s each \cite{kwon2022multi}, and then an attention statistic pooling layer was used to project the variable frame-level features to the utterance-level fixed-length vector.
Consequently, the MFA-Conformer provided a 192-dimensional speaker embedding vector.

The model was trained with a batch size of 200 using the AdamW optimizer with a weight decay of 1e-7.
The learning rate was scheduled via cosine annealing with warmup restart \cite{loshchilov2016sgdr} with a cycle size of 25 epochs, maximum learning rate of 1e-3, and decay rate of 0.5 per cycle.
As an objective function, we combined the AP objective function \cite{deng2019arcface} with two utterances for the prototype and the AAM-Softmax objective function \cite{chung2020defence} with a margin of 0.3 and scale of 30.

\begin{table}[t]
  \caption{Comparison of precision and F1-score of different overlapped speech detection modules}
  \label{tab:ovd}
  \centering
  \begin{tabular}[c]{lcc}
    \toprule
     & \textbf{Precision (\%)} & \textbf{F1} (\%) \\
    \midrule
    1. ResNet+LSTM (S = 1)       & 68.55  & 68.22 \\
    2. ResNet+LSTM (S = 2)       & 67.55  & 67.40 \\
    3. SincNet+LSTM           & 68.83  & 66.79 \\
    Fusion (1+2)               & 83.94  & 56.99  \\
    Fusion (1+2+3)             & 88.81  & 52.45 \\
    \bottomrule
  \end{tabular}
\end{table}

    \subsection{Clustering}\label{sec:clustering}
\begin{table*}[tp]
  \caption{Performance comparison of our different versions of speaker diarization systems shown in Fig. \ref{fig:arch}}

  \label{tab:result}
  \centering
  \begin{tabular}[c]{cccccccccc}
    \toprule
     \multirow{3}*{\textbf{System}} & \multirow{2}*{\textbf{Time-scale}} & \multirow{2}*{\textbf{Speech}} &  \multicolumn{2}{c}{\textbf{VAL46}} & \multicolumn{2}{c}{\textbf{VoxSRC22 test set}}\\
      \cmidrule(lr){4-5} \cmidrule(lr){6-7} \cmidrule(lr){8-9} 
     
      & (Segment / hop length) & \textbf{Enhancement} & \textbf{DER (\%)}  & \textbf{JER (\%)} & \textbf{DER (\%)}  & \textbf{JER (\%)} &\\
 
   \midrule
   \centering
	1 & 1s / 0.75s       & no      & 4.41 & 27.47 & -    & -  \\
	2 & 2s / 1s      & no      & 3.97 & 27.45 & -    & -  \\
	3 & 3s / 1.5s      & no      & 4.02 & 26.92 & -    & -  \\
	4 & 2s / 0.25s      & yes     & 4.14 & 27.75 & -    & -  \\
    \midrule
    5 & \multicolumn{2}{c}{Dover-lap Fusion (1+2+3)}        & 3.66 & 26.63 & -    & -  \\
    6 & \multicolumn{2}{c}{Dover-lap Fusion (1+2+3+4)}      & 3.56 & 27.63 & 5.12 & 30.82  \\
     \bottomrule
     \end{tabular}
\end{table*}
        To cluster speaker embedding vectors obtained from MFA-Conformer, we first performed dimension reduction from 192 to 128 using linear discriminant analysis (LDA).
Then, two probabilistic LDA (PLDA) \cite{PLDA} models were estimated using the reduced dimensional embedding vectors extracted from voxceleb1 \& 2 and voxconverse DEV402 set each.
The final PLDA  model was obtained by linearly interpolating two PLDA models with a factor of 0.9.
Next, for a given utterance, the agglomerative hierarchical clustering (AHC) with PLDA scores was performed to assign a cluster index to each frame.
The AHC clustering threshold was set using a two component Gaussian mixture model with a shared variance, estimated from all the PLDA scores in the utterance.

The clusters from AHC were further processed using the short-duration filter \cite{xiao2021microsoft}.
In other words, short-duration clusters were identified using a threshold of 2.5 s, and then it was decided whether each short-duration cluster should be merged into the closest long cluster or treated as a new cluster according to the cosine distance threshold of 0.5.
Notice that a higher threshold value this work than those in previous works \cite{xiao2021microsoft, wang2021dku, wang2021bytedance} caused slight underclustering.
However, this underclustering was remedied using variational Bayesian (VB)-HMM-based clustering \cite{landini2020but_DIHARDII, landini2021analysis_voxSRC20}.
VB-HMM aims at reassigning a cluster index to each frame by considering the time dependencies with a proper number of clusters.
In fact, the parameters of VB-HMM need to be optimized according to each input structure.
In particular, we experimented with three different input structures according to different pairs of (segment length, hop size), e.g., (1 s, 0.75 s), (2 s, 1 s), (3 s, 1.5 s), and (2 s, 0.25 s).
The first three pairs were applied when a raw speech signal was used, while the last one was applied when enhanced speech was used.

    \subsection{Overlapped Speech Detection}\label{sec:osd}
        The system of VAD and OSD are based on  ResNet+LSTM and SincNet+LSTM models.
ResNet+LSTM is trained from scratch using a speech enhanced dataset.
In contrast, SincNet+LSTM is transferred from pretrained using a raw dataset.

To detect overlapped speech frames, we again used a VAD module whose target label was overlapped speech `1' or no overlapped speech `0'.
We trained three VAD modules: two ResNet+LSTM modules with $S=\{1, 2\}$ and one SincNet+LSTM module.
In particular, weighted BCE loss was used to train for OSD owing to the data imbalance problem.
This problem occurs because the number of overlapped speech frames is much smaller than that of single speech or silence frames.
Except this loss function, the entire training procedure was identical to that described in Section \ref{sec:vad}. The posterior output from each module was compared with a threshold to detect overlapped speech frames, where the threshold was intentionally set so that the precision became high.

Table \ref{tab:ovd} compares the precision and F1 score of the overlapped speech detection modules. 
As shown in the table, Fusion (1+2+3) provided the best precision among other single modules and the fusion module.
Note that Fusion (1+2+3) indicates that the posterior outputs from the single modules are averaged.
Consequently, we selected this Fusion (1+2+3) as the OSD module.

\section{Dataset Description} The dataset given by the Challenge is divided into a development set (DEV402) and a validation set (VAL46), which are the first 402 recordings and the last 46 recording, respectively, as in \cite{wang2021dku}.
The datasets used for each module in our developed SD systems are as follows:

\begin{itemize}
    \item VAD and OSD: 
        for scratch training of ResNet+LSTM, AMI  \cite{carletta2005ami}, AISHELL-4 \cite{fu2021aishell}, CALLHOME  \cite{callhome}, DIHARD I  \cite{ryant2018first}, and DIHARD II \cite{ryant2019second} are the mixed training sets that are processed by a speech enhancement module. DEV402 and VAL46 are also used for fine-tuning and validation, respectively. In addition, only DEV402 without speech enhancement is used for transfer learning of SincNet+LSTM from the pretrained model.
    \item Speaker embedding: 
        voxceleb 1 \& 2 development sets  \cite{Nagrani17,Chung18b} are used  for training the speaker embedding module.
    \item Data augmentation:  
        data augmentation is performed, in which utterances are convolved with one of the simulated room impulse responses (RIRs) \cite{RIRs} and mixed with noise signal from the MUSAN corpus \cite{MUSAN}, which have ambient, music, television, and babble noises. The data augmentation is employed for VAD, OSD, and speaker embedding module.
\end{itemize}

\section{Experimental Results} 
    Table \ref{tab:result} compares the SD performance of the different versions of our speaker diarization systems shown in Fig. \ref{fig:arch}.
The performance was evaluated on both VAL46 and VoxSRC 22 challenge test sets.
In this table, the first four single systems are different from the time scale and the presence or absence of speech enhancement. 

As shown in table, System 2 showed the best DER of 3.97\% on VAL46  among the single systems without speech enhancement.
However, we experimentally found that some utterances were well diarized when employing speech enhancement.
Even though System 4 had a higher DER than Systems 1–3 because of applying speech enhancement to all the utterances in VAL46, we fused it with Systems 1–4 using DOVER-Lap \cite{raj2021dover}, which resulted in the lowest DER.
Accordingly, we submitted this fusion system for the Challenge and achieved a DER of 5.12\% on the challenge test set, which ranked 3rd at the 2022 VoxSRC .

\section{Conclusion} 
    In this report, we described our submitted SD system for the diarization task (Track 4) of the 2022 VoxSRC.
To achieve better performance, we mainly focused on using multi-scaled x-vectors by applying MFA-Conformer to speaker embedding.
We evaluated the proposed system with the dataset provided by the 2022 VoxSRC Challenge and achieved DERs of 3.56\% and 5.12\% on the development set and challenge test set, respectively, which ranked 3rd at the 2022 VoxSRC.

\section{Acknowledgements} 
    This research was supported by Culture, Sports, and Tourism R\&D Program through the Korea Creative Content Agency grant funded by the Ministry of Culture, Sports, and Tourism (R2022060001) in 2022, by Institute of Information \& Communications Technology Planning \& Evaluation (IITP) grant funded by the Korea government (MSIT) (No. 20220-00963), and by the Korean National Police Agency [Pol-Bot Development for Conversational Police Knowledge Services / PR09-01-000-20].

\bibliographystyle{IEEEtran}
\bibliography{main}

\end{document}